\documentclass[aps,prd,twocolumn]{revtex4}
\newcommand{\be}{\begin{equation}}
\newcommand{\ee}{\end{equation}}
\newcommand{\bea}{\begin{eqnarray}}
\newcommand{\eea}{\end{eqnarray}}
\newcommand{\bes}{\begin{eqnarray}}
\newcommand{\ees}{\end{eqnarray}}
\newcommand{\no}{\noindent}
\newcommand{\dsla}{\partial\hspace{-5.7pt} \slash  }

\begin{document}


\title{Uniting Gross-Neveu and Massive Schwinger Models}


\author{Tongu\c{c} Rador}
\email[e-mail: ]{rador@gursey.gov.tr}
\affiliation{Feza G\"ursey Institute\\ Emek Mah. No:68,
\c{C}engelk\"oy 81220 \\ Istanbul, Turkey}


\date{\today}

\begin{abstract}
We show that it is possible to obtain the Gross-Neveu model in 1+1
dimensions from gauge fields only. This is reminiscent of the fact
that in 1+1 dimensions the gauge field tensor is essentially a
pseudo-scalar. We also show that it is possible in this context to
combine the Gross-Neveu model with the massive Schwinger model in the
limit where the fermion mass is larger than the electric charge.
\end{abstract}

\pacs{}

\maketitle

\section{Introduction}
Exactly solvable models in 1+1 dimensions form a good theoretical 
laboratory due the possibility 
that they might provide ways to study and understand aspects of
their comparably more complicated 3+1 dimensional cousins. 
In this paper we are concerned about two
of these; the (massive) Schwinger (SWM)\cite{Schwinger} and the
Gross-Neveu (GNM) \cite{Gross-Neveu} models. The former provides ways to understand the
anomaly-gauge invariance relation, confinement, screening and so on, 
where as the latter presents
spontaneous breaking of a discrete symmetry and all the interesting
phenomena that is associated with it such as dynamical mass
generation. Both these theories escape the Coleman \cite{Coleman} theorem in their own way:  SWM in that
the anomaly already explicitly breaks the continuous chiral symmetry
of the fermion field and the GNM in that the symmetry
that is spontaneously broken is discrete.
 
The common technique for solving the GNM is to introduce an
auxiliary scalar field with a Yukawa type coupling to the fermion, so
that using its equations of motion will yield the quartic fermion term
of GNM. On the other hand in 1+1 dimensions the electromagnetic tensor
is essentially a pseudo scalar and there is a possibility to add a
non-minimal coupling between fermions, the 2D counterpart of the
$g-2$ coupling in 4D, being essentially a term of Yukawa type. In this
paper we pursue this idea and show that it is indeed possible to get
GNM from gauge fields and finally combine it with the massive SWM.
\section{Gross-Neveu from gauge fields}

In complete analogy to the use of scalars for studying the GNM we
consider the following lagrangian with N fermion and one vector fields,

\be\label{eq:1}
{\mathcal{L}}_{1}=-\frac{1}{4}F_{\mu\nu}F^{\mu\nu} + i
\bar{\Psi}\dsla\Psi
+\frac{g}{2}\epsilon_{\mu\nu}F^{\mu\nu}\bar{\Psi}\Psi\; .
\ee

As it stands this lagrangian is not renormalizable if one insists on
using the vector field as the dynamical variable. The reason is that
the 1-loop correction to the $A_{\mu}$ propagator will be proportional
to $g^{2}N\epsilon^{\mu a}\epsilon^{\nu b} q_{a} q_{b}$ and there
are no counter-terms in (\ref{eq:1}) to compensate for the infinity that arises.
Thus we modify (with the inclusion of a gauge fixing term) the lagrangian as follows,

\be\label{eq:2}
{\mathcal{L}}_{2}={\mathcal{L}}_{1}+\frac{A}{8}\left(\epsilon_{\mu\nu}F^{\mu\nu}\right)^{2}-\frac{1}{2\xi}(\partial_{\mu}
A^{\mu})^{2}\;\;.
\ee

We argue as follows for the  omission at this stage of the minimal
coupling, $e A_{\mu}\bar{\Psi}\gamma^{\mu}\Psi$. The
lagrangian (\ref{eq:2}) is invariant under the discrete symmetry $S$
defined to be

\be\label{eq:3}
S\equiv P\times \left\{ \Psi\to\gamma^{5}\Psi \;\;\protect{\rm and}\;\; A_{\mu}\to
A_{\mu}\right\}\;\; ,
\ee

\no where $P$ stands for the usual parity transformation. Now if one
includes the minimal coupling 
this symmetry will be explicitly
broken by the $U(1)$ anomaly. However as we will show shortly the lagrangian
(\ref{eq:3}) will exhibit spontaneous breaking (SSB) of the $S$ symmetry.
Thus ignoring  the explicit breaking term is plausible if its scale
is much smaller than the scale of the SSB
, which, in the present case can be identified with the fermion 
mass $M_{f}$ that would arise from SSB. 
So we have from the outset $e\ll \sim M_{f}$ and we can turn on the minimal
coupling after the SSB occurs. More on this
in section \ref{sec:3} and \ref{sec:4}.

Varying the action with respect to $A_{\mu}$ (and ignoring for now the
gauge fixing term) we get

\be
\partial_{\mu}\left[-F^{\mu\nu}+\frac{A}{2}\epsilon^{\mu\nu}(\epsilon_{ab}F^{ab})+g\epsilon^{\mu\nu}\bar{\Psi}\Psi\right]=0\;\;.
\ee 

If we replace $F_{\mu\nu}=\epsilon_{\mu\nu}\sigma$ we get
$\sigma=g\bar{\Psi}\Psi/(1+A)$ which gives the following lagrangian 

\be
{\mathcal{L}}_{GN}=i\bar{\Psi}\dsla\Psi-\frac{g^{2}}{2(1+A)}\left( \bar{\Psi}\Psi\right)^{2}\;\;,
\ee

which means that to have a correspondence to the GNM at all we have to demand

\bea
G^{2}&\equiv&-\frac{g^{2}}{1+A}>0 \;\;, \\
&\protect{\rm that \;\; is:}& \nonumber \\
1+A &<& 0 \;\;.
\eea

To proceed further we need the free photon
propagator which is given by the terms of the lagrangian quadratic in $A_{\mu}$,

\bea{\label{eq:7}}
P^{\mu\nu}&=&\frac{-i}{q^{2}}\left[ \right. 
g^{\mu\nu} -(1-\frac{1}{\xi})\frac{q^{\mu}q^{\nu}}{q^{2}} \nonumber \\
&&
\;\;\;\;\;\;\;\;+\left.
\left(\frac{A}{1+A}\right)\frac{{\bar{q}}^{\mu}{\bar{q}}^{\nu}}{q^{2}}\right]
\;\; .
\eea

Here we have defined

\be
\bar{q}^{\mu}\equiv\epsilon^{\mu\nu}q_{\nu}\;\;,
\ee

Now, as in the GNM, we invoke the large $N$ argument. That is we let
$N\to\infty$ keeping $g^{2}N$ finite. Then the only $O(N^{-0})$
contributions come from the fermion loops. Since there will be an
ultraviolet infinity that will arise from a fermion loop and since it
will have the tensor form $\bar{q}^{\mu}\bar{q}^{\nu}$ the infinity
will be gotten rid of by renormalizing $A$. A simple calculation using
dimensional analysis and and $\overline{MS}$ subtraction scheme yields the
following

\bea\label{eq:tachyon1}
P^{\mu\nu}&=&\frac{-i}{q^{2}}\left[\right.
g^{\mu\nu}-(1-\frac{1}{\xi})\frac{q^{\mu}q^{\nu}}{q^{2}}\nonumber \\
&&
\;\;\;\;\;\;\;\;+\left.
\left(\frac{B}{1+B}\right)\frac{{\bar{q}}^{\mu}{\bar{q}}^{\nu}}{q^{q}}\right]
\;\; , 
\eea

with

\bea
1+B &=&
1+A(\mu^{2})-\frac{g^{2}N}{2\pi}\ln(\frac{-q^{2}}{\mu^{2}})\;\; ,\\
\mu \frac{\partial A}{\partial \mu}&\equiv& \beta(A) =
-\frac{g^{2}N}{\pi}\;\; .
\eea

From the last equation it is easy to see that the theory is
asymptotically free

\be
\mu\frac{\partial G^{2}}{\partial
\mu}=-\frac{g^{2}N}{\pi}\frac{g^{2}}{(1+A)^{2}}\;\; .
\ee

Clearly the propagator (\ref{eq:tachyon1}) has tachyonic poles and following
the common wisdom there should be two meanings for this. Either the
theory does not make sense at all or we are simply expanding about the
wrong vacuum. We will now show that the latter is the case. To study a
possible SSB it is enough to consider
$F_{\mu\nu}=\epsilon_{\mu\nu}\sigma$ and study the theory around a
constant $\sigma_{cl}$ background field. The procedure is exactly the
same as in the GNM \cite{Gross-Neveu,ColemanWeinberg}. Thus the renormalized effective potential is

\be{\label{eq:10}}
V=-\frac{1+A}{2} \sigma_{cl}^{2} + \frac{g^{2} N}{4 \pi}
\sigma_{cl}^{2} \left[
\ln(\frac{\sigma_{cl}^{2}}{\sigma_{0}^{2}})-3\right] \;\; .
\ee

Where we have introduced the non-zero subtraction field strength
$\sigma_{0}$ (also an $\overline{MS}$ quantity) in such a way that $V''(\sigma_{0})=-(1+A)>0$. Since to the
order we are working at there is no wave function renormalization the
effective potential obeys the following renormalization group equation

\be
\left[
\sigma_{0}\frac{\partial}{\partial\sigma_{0}}+\tilde{\beta}(A)\frac{\partial}{\partial
A}\right] V=0\;\;.
\ee

From here we
find $\tilde{\beta}(A)=\beta(A)$ meaning
$\mu=\text{const.}\times\sigma_{0}$. 

The potential (\ref{eq:10}) has two
minima that are images of each other with respect to the symmetry $S$;
 
\bea{\label{eq:11}}
\sigma_{M_{f}}&=&\pm\sigma_{0}\exp\left[1+(1+A)\pi/g^{2}N\right]\;\; , \\
V''(\sigma_{M_{f}})&=&\frac{g^{2}N}{\pi}\;.
\eea

This signals the spontaneous breaking of $S$ and consequently the
fermion acquires a mass

\be
M_{f}=g\,|\sigma_{M_{f}}|\;\; .
\ee

This mass being a physical quantity obeys the same
renormalization group equation as the effective potential. 

We can now compute the photon propagator in the broken symmetry phase
of the theory. The renormalization condition we have employed for the
effective potential means that 

\be
\left[\bar{q}_{\mu}\bar{q}_{\nu}\Pi^{\mu\nu}(q^{2},M^{2}_{f})\right]^{-1}_{q^{2}=0}=iV''(\sigma_{M_{f}})=\frac{g^{2}N}{\pi}\;\;
.
\ee

Thus we have to calculate the loop with a non-zero fermion mass and we
subtract at zero momentum. The above condition yields $\mu=g \sigma_{0}$
and we get the propagator in the broken phase to be

\bea
\Pi^{\mu\nu}&=&\frac{-i}{q^{2}}\left[g^{\mu\nu}+\left(\frac{1+C}{C}\right)\frac{\bar{q}^{\mu}\bar{q}^{\nu}}{q^{2}}\right]\;\;,\\
C&=&-\frac{g^{2}N}{2\pi}f(q^{2}/4M_{f}^{2})\;\;,\\
f(x)&=&2\sqrt{\frac{1-x}{x}}\tan^{-1}\left[\sqrt{\frac{x}{1-x}}\;\right]\;\;.
\eea

We see that now there is a physical pole that appears at the threshold
$4 M^{2}_{f}$ and the tachyon has disappeared. This completes the full correspondence to the GNM
because the 4-fermi amplitudes will come to be exactly the same. For
example the $e^{+}e^{-}\to e^{+}e^{-}$  scattering amplitude will be

\be
\frac{2\pi i}{N}\left[
\frac{1}{f(s/4M_{f}^{2})}+\frac{1}{f(u/4M_{f}^{2})}\right]\;\;.
\ee

\section{Turning on the minimal coupling}\label{sec:3}

It is better to study the theory with the minimal coupling turned on
(that is now the fermion has charge) in the unbroken phase to get a
better feeling about the SSB. 
When $e\neq 0$ there is another diagram that contributes to the
photon propagator, but now with a different tensor structure;

\be
i\frac{e^{2}N}{\pi}\left[g^{\mu\nu}-\frac{q^{\mu}q^{\nu}}{q^{2}}\right]\;\;.
\ee

Here we again resort to large $N$ argument. That is we let
$N\to\infty$ keeping $e^{2}N$ finite. As we did before summing all the
1PI graphs, the exact propagator becomes,

\bea
\Pi^{\mu\nu}&=&\frac{-i}{(q^{2}-\frac{e^{2}N}{\pi})}\left[
g^{\mu\nu}-\frac{q^{\mu}q^{\nu}}{q^{2}}+\right. \nonumber \\
&& \;\;\;\;\;\;+ \left.\left(\frac{B}{1+B-e^{2}N/(\pi
q^{2})}\right)\frac{{\bar{q}}^{\mu}{\bar{q}}^{\nu}}{q^{q}}\right]\;\; .
\eea

There will be tachyonic poles in this propagator if

\be
1+A(\mu^{2})-\frac{g^{2}N}{2\pi}\ln(\frac{-q^{2}}{\mu^{2}})-\frac{e^{2}N}{q^{2}\pi}=0 \;\;.
\ee

It can be shown that the tachyons exist for all values of $e$ and
$g$. So we see that ignoring the minimal term before SSB is verified since this
interaction term does not make the situation any better.

We now turn back to the broken phase. In this situation the
contribution from the minimal coupling term changes due to the finite
fermion mass. The full interacting propagator is given by;

\bea
\Pi^{\mu\nu}&=&-iX(q^{2})(g^{\mu\nu}-\frac{q^{\mu}q^{\nu}}{q^{2}})-iY(q^{2})\frac{\bar{q}^{\mu}\bar{q}^{\nu}}{q^{2}}\\ \nonumber
&&-i\xi\frac{q^{\mu}q^{\nu}}{q^{4}})\;\;.
\eea

\noindent with;

\bea
X(q^{2})&=&\frac{1}{q^{2} (1-\kappa F(z))} \;\; ,\\
Y(q^{2})&=&\frac{1}{q^{2} (1-\kappa F(z))} \frac{1+\lambda G(z)}{\lambda G(z)+\kappa F(z)}\;\; .
\eea

\noindent The functions $F$ and $G$ are;

\bea
G(z)&=&\sqrt{\frac{1-z}{z}}\tan^{-1}(\sqrt{\frac{z}{1-z}}) \;\;,\\
F(z)&=&\frac{1}{z}\left[1-\frac{1}{\sqrt{z(1-z)}}\tan^{-1}(\sqrt{\frac{z}{1-z}})\right] \;\;,
\eea

\noindent and we adopted the following for convenience in notation

\bea
z&\equiv& q^{2}/4M^{2}_{f}\;\;, \\
\lambda&\equiv& g^{2}N/\pi \;\;,\\
\kappa&\equiv& e^{2}N/4\pi M^{2}_{f}\;\;. 
\eea

The functions $X$ and $Y$ have poles at $q^{2}=0$ and 
$q^{2}=4M^{2}_{f}-(e^{2}N)/(g^{2} N)$  for $e\ll 2gM_{f}$. However the pole
at $q^{2}=0$ is spurious and should not show in the physical scattering 
amplitudes. Thus we see that the pole that was at the threshold in the
absence of the minimal coupling moves toward $q^{2}=0$. A further
excursion in this direction defines a critical point where the pole
appears at $q^{2}=0$ just before disappearing, 

\be
e^{2}_{c} N =6M^{2}_{f} g^{2} N \;\;.
\ee

The theory is tachyon free.

\section{The anomaly}\label{sec:4}

The careful reader might have already noticed that ignoring the
anomaly term at the beginning as an explicit symmetry breaking term
might not be plausible since ,in principle, we are working with an 
infinite number
of fermion fields. The discrete symmetry $\Psi\to\gamma_{5}\Psi$ will
yield the following term when $e\neq 0$ ,

\be
\frac{e N}{2\pi} \frac{\pi}{2}\epsilon_{\mu\nu}F^{\mu\nu}\;\;.
\ee

Here $\pi/2$ is the angle we should use to cast the  discrete chiral
transformation in $S$ as part of an axial $U(1)$ symmetry. Now, we 
cannot demand the term above to be finite if 
we keep $e^{2}N$ finite using
the large $N$ argument. This difficulty can be remedied however by
enriching the flavor structure of the theory. If we have a flavor
structure the anomaly term will become

\be
\sum_{i} \frac{e_{i} N_{i}}{2\pi}\alpha_{i}\epsilon_{\mu\nu}F^{\mu\nu}\;\;.
\ee

This sum can be made finite and smaller (even in the case of infinte
number of ``total'' fermion fields) than $M_{F}$ by a suitable
choice of the parameters. Obviously the SSB part will
remain uneffected by this and the results of the previous sections
will still hold.
\section{Conclusion}

In this work we have shown that it is possible to achieve Gross-Neveu
model from gauge fields only. In the limit where the fermion charge
vanishes exactly the correspondace with the Gross-Neveu model is one-to-one.
That is there is spontaneous breaking of a discrete chiral symmetry
and dynamical mass generation. We argued that the turning on of electric
charge and consequently the explicit breaking of the mentioned
symmetry by the anomaly can be controlled by extending the flavor
structure of the theory. Then, it is possible to treat the anomaly
term as a perturbing explicit symmetry breaking term and we see that the
fermion-antifermion bound state mass is lowered by an amount
proportional to the ratio of the two scales of symmetry breaking.

It would be interesting to test the
conclusions about  the model
proposed in this brief report 
on the lattice. A joint effort on this is in progress \cite{11}.

\begin{acknowledgments} 
I benefited from discussions with C. Hoelbling, A. Kaya,
C. Sa\c{c}l{\i}o\~{g}lu and T. Turgut. I also am thankful to T. Turgut
for carefully reading  the manuscript.

\end{acknowledgments}


\end{document}